\documentclass[12pt,english]{article}
\usepackage[latin1]{inputenc}
\usepackage{color}
\makeatletter
\usepackage{amsmath,amssymb}
\newcommand{\be}{\begin{equation}}
\newcommand{\ee}{\end{equation}}
\newcommand{\bea}{\begin{eqnarray}}
\newcommand{\eea}{\end{eqnarray}}
\newcommand{\nn}{\nonumber \\}
\newcommand{\p}[1]{(\ref{#1})}
\newcommand{\lb}{\label}

\newcommand{\bz}{{\bar z}}
\newcommand{\bp}{{\bar \psi}}

\def\6{\partial}
\def\7{\tilde}
\def\8{\widehat}

\def\G11{\Gamma_{11} }

\usepackage{babel}
\makeatother
\begin{document}

\begin{titlepage}
\rightline{ UMTG-12,${\;}$ JINR-E2-2010-25}

\vfill

\begin{center}
\baselineskip=16pt {\Large\bf  Generalized ${\cal N}=2$ Super Landau Models}
\vskip 0.3cm
{\large {\sl }}
\vskip 10.mm
{\bf Andrey Beylin$^{*,1}$, Thomas Curtright$^{*,2}$, ~Evgeny Ivanov$^{\dagger,3}$

and~ Luca Mezincescu$^{*,4}$ }
\vskip 1cm
{\small
$^*$
Department of Physics, University of Miami,\\
Coral Gables, FL 33124, USA\\
}
\vspace{6pt}
{\small
$^\dagger$ Bogoliubov Laboratory of Theoretical Physics, JINR, \\
141980 Dubna, Moscow Region, Russia\\
}
\end{center}
\vfill

\par
\begin{center}
{\bf ABSTRACT}
\end{center}
\begin{quote}

We generalize  previous results for the superplane Landau model to exhibit
an explicit worldline ${\cal N}=2$ supersymmetry
for an arbitrary magnetic field on any two-dimensional manifold.
Starting from an off-shell  ${\cal N}=2$ superfield formalism, we  discuss the quantization procedure
in the general case characterized by two independent
potentials on the manifold and show that the relevant Hamiltonians are factorizable.
In the restricted case, when both the Gauss curvature and the magnetic field are constant over
the manifold and, as a consequence, the underlying potentials are related, the Hamiltonians
admit infinite series of factorization chains implying the integrability of the associated systems.
We explicitly determine the spectrum and eigenvectors for the particular model with $\mathbb{CP}^1$
as the bosonic manifold.

\vfill
\vfill
\vfill
\vfill
\vfill
\hrule width 5.cm
\vskip 2.mm
{\small
\noindent $^1$ beylin@server.physics.miami.edu\\
\noindent $^2$  curtright@miami.physics.edu\\
\noindent $^3$ eivanov@theor.jinr.ru\\
\noindent $^4$ mezincescu@server.physics.miami.edu\\}
\end{quote}
\end{titlepage}

\setcounter{page}{1}
\section{Introduction}
The original Landau model (LM) \cite{Landau1930} describes a charged
particle moving on a plane under the influence of a uniform
magnetic field orthogonal to the plane. This model can be viewed
as a Wess-Zumino-type sigma model on the coset space ${\cal G} /  {\cal H}$, where $\cal G$ is the
Heisenberg group generated by
\bea
\left[ P, P^\dagger \right]= 2\kappa,
\eea
and the subgroup $\cal H$ is generated by the central charge ${\cal Z} \equiv 2\kappa$. In our previous papers \cite{IMT3,CIMT}
this model has been generalized to a superplane by considering a one-dimensional
sigma model on the super coset $ISU(1|1) / U(1|1) \times {\cal Z}$. The relevant invariant Lagrangian involves
non-standard second-order kinetic terms for the odd variables.
According to the standard lore \cite{VoPa}, such a theory should contain ghosts.
However, the quantum Hamiltonian for the superplane LM was found to have real eigenvalues and a complete set of
eigenvectors. The naive inner product was
redefined to yield positive-definite norms, after which a normal quantum system emerged.
As a bonus, this redefinition procedure allowed one to uncover additional integrals
of motion which generated a hidden world-line ${\cal N} = 2$ supersymmetry
of the  model.\footnote{The presence of hidden ${\cal N} = 2$ supersymmetry
in the superplane model was also noticed in \cite{Hasebe}.}

The superplane LM can be viewed as a
contraction of the supersphere LM associated with the supercoset $SU(2|1)/U(1|1)$.
In ref. \cite{ACIMT} we investigated  the
properties of the supersphere model. We found that it can be consistently defined, but its symmetry structure is
rather complicated. With the exception of the ground state, the model
reveals a dynamical $SU(2|2)$ symmetry.  Therefore it is likely that this
model does not inherit the world-line ${\cal N} = 2$ supersymmetry of the
superplanar LM.

There arises the question as to whether
the hidden ${\cal N}=2$ supersymmetry of the planar model could be
lifted to more general target spaces. This question is addressed in the present paper.
We take ${\cal N}=2$ world-line supersymmetry as the primary principle and use a superfield formulation developed
in \cite{I} to lift the  superplane LM to a general supermanifold  with a rather general
magnetic-type coupling. Then we develop a quantization
procedure for the relevant component model.

In Section 2 we describe an off-shell superfield formulation of the superplane LM.
In Section 3 we lift this model to a general supermanifold and show that the superfield Lagrangian
is specified by two independent superpotentials which, after passing to the component formulation, yield
a non-trivial target supermetric, as well as couplings to an external magnetic field. We analyze the component structure of the corresponding
Lagrangian, develop the Hamiltonian formalism,  and exhibit further covariances and invariances of the model.
In Section 4 we compute the Noether charges generating the ${\cal N} = 2$ supersymmetry
algebra and work out their semiclassical closure. We outline problems encountered while quantizing this system.
Then we quantize and analyze the system using a quantization procedure based on an inner product with a trivial
integration measure, demanding the ${\cal N} = 2$ supersymmetry in the process. We emphasize the necessity
to redefine the naive inner product \cite{Bend} (see also \cite{Bi}), in order to acquire the positivity of the norm.
We  show that the quantum model on a general supermanifold with rather general magnetic-type coupling is equivalent
to two eigenvalue problems with factorized positive definite Hamiltonians.
In Section 5 we develop an alternative method of quantization of this model by using  semi-classically equivalent
supercharges which upon quantization require non-trivial measures in the inner product. We show that this quantization procedure
is equivalent to the one developed in Section 4. In Section 6, based on reasoning of ref. \cite{FV},
we argue that, when the Gauss curvature of the two-dimensional even manifold is a constant and the ``magnetic field''
coupling is induced by the corresponding K\"ahler connection, an infinite sequence of eigenvalues and eigenvectors (forming a complete set)
can be derived within the renowned Schr\"{o}dinger factorization method. In this restricted case two original
target space potentials (coming from the superfield superpotentials) are reduced to a single K\"ahler potential.
In Section 7, as an instructive example, we study in detail the appropriate superextended $\mathbb{CP}^1$ model.
We conclude with a summary and outlook  in Section 8.

\setcounter{equation}{0}
\section{Worldline ${\cal N}=2$ supersymmetry made manifest}
In this Section we reformulate the superplane model
of ref. \cite{IMT3,CIMT}  within a worldline ${\cal N}{=}2$ superfield approach, following ref. \cite{I}.

We start with the necessary definitions.
The basic objects are two
${\cal N}{=}2,\, d{=}1$ chiral bosonic and fermionic
superfields $\Phi$ and $\Psi$ of the same dimension.

The real ${\cal N}{=}2,\, d{=}1$ superspace is parametrized as:
\be
(\tau, \theta, \bar\theta)\,. \lb{1}
\ee
The left and right chiral superspaces are defined by
\be
(t_L, \theta), \quad (t_R, \bar\theta), \quad t_L = \tau -i\theta\bar\theta, \; t_R
= \tau +i\theta\bar\theta = t_L + 2i\theta\bar\theta\,. \lb{2}
\ee
It will be convenient to work in the left (chiral) basis, so for brevity we will use
the notation $t_L \equiv t,\; t_R = t + 2i\theta\bar\theta$. In this basis, the
${\cal N}{=}2$ covariant derivatives are defined by
\be
\bar D = -\frac{\partial}{\partial \bar\theta}\,, \; D =\frac{\partial}{\partial \theta}
- 2i \bar\theta \partial_t\,, \; \{D, \bar D\} = 2i\partial_t\,, \; D^2 = \bar D^2 = 0\,. \lb{3}
\ee

The chiral superfields $\Phi$ and $\Psi$ obey the conditions
\be \bar D \Phi = \bar D\Psi = 0
\lb{Chir}
\ee
and in the left-chiral basis have the following
component field contents:
\be \Phi (t, \theta) = z(t) + \theta
\chi(t)\,, \;\; \Psi(t, \theta) = \psi(t) + \theta h(t)\,,
\lb{4}
\ee
where the complex fields $z(t), h(t)$ are bosonic and
$\chi(t), \psi(t)$ are fermionic. The conjugated superfields, in
the same left-chiral basis, have the following $\theta$-expansions:
\be \bar \Phi = \bz -\bar\theta \bar\chi + 2i
\theta\bar\theta \dot{\bz}\,, \; \bar\Psi = \bp +
\bar\theta \bar h  + 2i\theta\bar\theta \dot{\bp}\,. \lb{5}
\ee
Also, we shall need the component structure of the following superfields:
\bea
D\Phi &=& \chi -2i \bar\theta \dot{z} + 2i
\theta\bar\theta \dot\chi\,, \; \bar D\bar\Phi = (D\Phi)^\dagger
= \bar\chi + 2i\theta \dot{\bz}\,, \nn D\Psi &=& h
-2i\bar\theta \dot\psi + 2i\theta\bar\theta \dot h\,, \; \bar
D\bar\Psi = -(D\Psi)^\dagger = -\bar h  + 2i\theta
\dot{\bp}\,. \lb{6}
\eea

The superfield action yielding in components the superplane
model action of ref. \cite{IMT3,CIMT,Hasebe} is as follows:
\be
S = - \int dtd^2\theta \left\{\Phi\bar\Phi + \Psi\bar\Psi + \rho\left[\Phi D\Psi -
\bar\Phi \bar D\bar\Psi\right]\right\} \equiv \int dt \left\{{\cal L}_1 +
\rho{\cal L}_2\right\}\,.\lb{7}
\ee
Here $\rho$ is a real parameter. The Berezin integral is normalized as
\be \int d^2\theta (\theta\bar\theta) = 1\,.
\ee
\vspace{0.3cm}After doing the Berezin integral, we find
\bea
&&{\cal L}_1 \;\Rightarrow \; -2i\left(z\dot{\bz} + \psi\dot{\bp}\right)
-\left(\chi\bar\chi + h\bar h \right)\,, \nn
&&{\cal L}_2 \;\Rightarrow \; -2i\left(z \dot h + \chi \dot\psi + \dot{\bz} \bar h +
\bar\chi \dot{\bp}\right). \lb{8}
\eea
The fields $h$ and $\chi$ are auxiliary and they can be eliminated by their equations
of motion
\be
\chi = 2i\rho\, \dot{\bp}\,, \quad h = -2i \rho\, \dot{\bz}\,. \lb{9}
\ee
Upon substituting this into the sum ${\cal L} \equiv {\cal L}_1 + \rho{\cal L}_2$, the latter
becomes
\be
{\cal L} \;\Rightarrow \; -2i\left(z\dot{\bz} + \psi\dot{\bp}\right) +
4\rho^2 \left(\dot z \dot{\bz} + \dot{\bp}\dot\psi \right). \lb{10}
\ee
After redefining
\be
\bp = \zeta\,, \; \psi = \bar\zeta\,, \quad 4 \rho^2 \equiv \frac{1}{\kappa}\,,\lb{RL}
\ee
and integrating by parts, the Lagrangian \p{10} takes the form
\be
{\cal L} = -i\left(z \dot{\bz} - \bz \dot z + \zeta\dot{\bar\zeta} - \dot\zeta\bar\zeta\right)
+ \frac{1}{\kappa}\left(\dot z \dot{\bz} + \dot\zeta\dot{\bar\zeta} \right). \lb{11}
\ee
That is  the superplane model component Lagrangian \cite{IMT3,CIMT}, modulo reversing of the time,
$t \rightarrow -t$, and the overall factor $\kappa$

By construction, the superfield action \p{7} is manifestly ${\cal N}{=}2$
supersymmetric. The ${\cal N}{=}2$ transformations of the component fields can be found from
\be
\delta \Phi = -\left[\epsilon Q - \bar\epsilon \bar Q\right]\Phi\,, \quad
\delta \Psi = -\left[\epsilon Q - \bar\epsilon \bar Q\right]\Psi\,, \lb{12}
\ee
where, in the left-chiral basis,
\be
Q = \frac{\partial}{\partial \theta}\,, \quad \bar Q = - \frac{\partial}{\partial \bar\theta}
- 2i\theta \partial_t\,, \quad \{Q, \bar Q \} = -2i\partial_t = 2 P_0\,. \lb{13}
\ee
It follows from \p{12}, \p{13} that off shell
\be
\delta z = - \epsilon \chi\,, \quad \delta \chi = 2i \bar\epsilon \dot z\,, \quad
\delta \psi = -\epsilon h\,, \quad \delta h = 2i \bar\epsilon \dot\psi\,. \lb{14}
\ee
With the on-shell values \p{9} for the auxiliary fields
and with the relabelling \p{RL}, these transformations become
\be
\delta z = - \frac{i}{\sqrt{\kappa}}\, \epsilon \dot{\zeta}\,, \quad \delta \zeta = -
\frac{i}{\sqrt{\kappa}}\,\bar\epsilon \dot z\,.
\ee
These are basically the same transformation laws as those given in \cite{CIMT}
(up to rescaling of $\epsilon, \bar\epsilon $).

Besides ${\cal N}{=}2$ supersymmetry, the superplane model also possesses
the target $ISU(1|1)$ symmetry. Its superfield realization was presented
in \cite{I}. In what follows we shall not need it.

\setcounter{equation}{0}
\section{From the free model to interaction}
We consider the following generalization
of the superfield superplane action \p{7}
\be
S = - \int dtd^2\theta \left\{K(\Phi,\bar\Phi) + V(\Phi, \bar\Phi)\Psi\bar\Psi + \rho\left[\Phi D\Psi  -
\bar\Phi \bar D\bar\Psi\right]\right\} \equiv \int dtd^2\theta\, {\cal L} \,.\lb{a}
\ee
Here, like in \p{7},  $\rho$ is a real parameter. In general, the potentials $K$ and $V$ are arbitrary
real functions of the chiral and antichiral scalar superfields $\Phi, \bar\Phi$. In principle, in the third term in \p{a}
we could also replace  $\Phi$ and $\bar\Phi$  by arbitrary mutually conjugate potentials. However,
in the case of generic dependence of such potentials on $\Phi$ and $\bar\Phi$,
the component action can be shown to be non-polynomial in the time derivatives of $z$ and $\bz$.
Such an exotic feature does not show up if these potentials are, respectively, holomorphic and
antiholomorphic. In this case the action can be reduced to the form \p{a} through a field redefinition.
Thus we take \p{a} as the starting point.\footnote{${\cal N}{=}2$ supersymmetry also admit superfield terms
$\sim \int dt d\theta \left[A(\Phi)\bar D\bar\Phi + B(\Phi)\Psi\right] + c.c.$. In the component action, they would induce
some new potential-like terms without derivatives, as well as a modification of other terms.
As far as we are interested in a generalization of the superplane model action \p{7}, we ignore this possibility.}

Our purpose  is to find a quantum formulation of this system. Also, we wish to learn  which $K$'s and $V$'s permit
the stationary Schr\"{o}dinger equation for this system to be solved, that is, in which case the eigenfunctions and eigenvalues of the
relevant Hamiltonian can be fully determined.

To begin, we rewrite the Lagrangian density  in terms of the
component fields:
\bea\label{L}
{\cal L}_{comp} &=& i\left(\dot{z}
K_z - \dot{\bar{z}} K_{\bz}\right) -\chi\bar\chi\,K_{z\bz} - i
V\left(\psi\dot{\bar{\psi}}- \dot{\psi} \bp\right) - h\bar h\,V
\nn && -\; \chi\psi \bar h\,V_z + \bar\chi\bp h\,V_{\bz} -
\chi\bar\chi\,\psi\bp\,V_{z\bz} + i \psi\bp\left(\dot{z} V_z
-\dot{\bar{z}} V_{\bz}\right) \nn
&& +2i\rho \left(\dot z h -
\dot{\bar{z}} \bar h - \chi\dot\psi - \bar\chi\dot{\bp}\right),
\lb{b}
\eea
where $K_z \equiv \partial_z K$, etc.\footnote{Herewith, the lower-case indices $z, \bar z, \psi, \bar \psi$ denote
derivatives, as well as mark the relevant momenta $P$ and connections ${\cal A}$ (see below).}
It is worthwhile to remark that (\ref{L}) is immediately put
in the Hamiltonian form,  since it
is linear in the time derivatives of the dynamical fields $z,
{\bar z}, \psi, {\bar \psi}$.   Indeed, redefining the auxiliary
fields $h, \chi, {\bar h}, {\bar \chi} $ as
\bea\label{P}
h& =&
{\frac{1}{2i\rho}}\left( P_z -iK_z - i \psi{\bar \psi}
V_{z}\right), \quad {\bar h} = -{\frac{1}{2i\rho}}\left( P_{\bar
z} + iK_{\bar z}  + i \psi{\bar \psi} V_{\bar z}\right),\nn
\chi
&=&{\frac{1}{2i\rho}}\left(P_\psi - i V {\bar \psi}\right), \quad
\quad \quad \quad \ \bar\chi = {\frac{1}{2i\rho}}\left(P_{\bar
\psi} - i V { \psi}\right),
\eea
(\ref{L}) can be cast in the form
\be
{\cal L}_{comp} = \dot Z^A P_A + \dot{\bar Z}^{\bar B} P_{\bar
B} - H_{class}\left( Z^A, P_C, {\bar Z}^{\bar B}, P_{\bar
C}\right),
\ee
where $A,B=(z,\psi)$, $\bar A, \bar B =
(\bz,\bar \psi)$, and
\be
P_A=\cfrac{\partial L}{\partial \dot
Z^A}, \quad P_{\bar B}=\cfrac{\partial L}{\partial \dot{\bar
Z}^{\bar B}}.
\ee
The classical Hamiltonian can now be expressed as
\be\label{H}
H_{class} =  {\cal P}_A\,g^{A\bar B}\,{\cal P}_{\bar
B}\,,
\ee
where we introduced the supermetric $g^{A\bar B}$ and
classical ``covariant derivatives''
\be
{\cal P}_A = P_A -{\cal
A}_A\,, \quad {\cal P}_{\bar A} = P_{\bar A} -{\cal A}_{\bar
A}\,.\lb{CovDer}
\ee
The entries of the supermetric are given by
\be\label{inversemetric'}
g^{z\bz}= {V\over 4\rho^2}\,,\quad
g^{\psi \bp}= -{1\over 4\rho^2}\left( K_{z\bz} + \psi
\bp\,V_{z\bz}\right)\,,\quad g^{z\bp}= -{V_{\bz}\bp\over
4\rho^2}\,,\quad g^{\psi \bz}= {V_z \psi\over 4\rho^2},
\ee
while the gauge superconnections are defined by
\be \label{u1conn}
{\cal A} =i \left(dZ^A \partial_A - d\bar Z^{\bar B}\partial_{\bar B}
\right) {\cal K} \equiv dZ^A {\cal A}_A + d\bar Z^{\bar B} {\cal
A}_{\bar B},
\ee
where
\be
{\cal K}= \left(K + \psi{\bar
\psi}\,V\right). \lb{CalK}
\ee
The explicit form of \p{u1conn} is:
\be\label{connections}
{\cal A}_z =i(K_z +\psi \bp\,V_z)\,, \;
{\cal A}_{\bz}=-i(K_{\bz} + \psi \bp\,V_{\bz})\,, \; {\cal
A}_{\psi} =  i V \bp\,, \; {\cal A}_{\bp} =i V \psi\,.
\ee
Though the connections have a nice K\"ahler form, the generic supermetric (see \p{metric} below)
cannot be expressed through ${\cal K}$ or any other K\"ahler-like
potential, so the supermanifold we deal with is not super K\"ahler
(as distinct e.g. from the supersphere \cite{ACIMT}).

Varying \p{L} with respect to the non-propagating fields $h, \bar
h, \chi, \bar\chi\,$, we obtain for them the following expressions:
\bea
&& h = -\chi \psi\,V^{-1}V_z - 2i\rho\,
V^{-1}\dot{\bar{z}}\,, \quad \bar h = \bar\chi \bp\,V^{-1}V_{\bz}
+ 2i\rho\, V^{-1}\dot{z}\,, \nn
&& \chi = 2i\rho\,A^{-1}\left[1
-\psi\bp\, BA^{-1}\right]\nabla\bp\,, \quad \bar\chi =
-2i\rho\,A^{-1}\left[1 -\psi\bp\, BA^{-1}\right]\nabla\psi\,,
\label{aux2}
\eea
where
\be\label{AB}
A \equiv K_{z\bz}\,, \quad B
\equiv V_{z\bz} - V^{-1} V_z V_{\bz}\,,
\ee
and
\be
\nabla\psi
\equiv \dot{\psi} + \dot{z}V^{-1}V_{z}\,\psi\,, \quad \nabla\bp
\equiv \dot{\bp} + \dot{\bar{z}}V^{-1}V_{\bz}\,\bp\,.
\ee
After substituting these expressions back into the off-shell Lagrangian
\p{b}, we obtain its on-shell form:
\bea
{\cal L}_{comp} &=&
4\rho^2\,V^{-1}\,\dot z\dot{\bar{z}} - 4\rho^2\,A^{-1}\left[1
-\psi\bp\, BA^{-1}\right] \nabla\psi \nabla\bp \nn &&
+\;i\left(\dot{z} K_z -\dot{\bar{z}} K_{\bz}\right) + i
\psi\bp\left(\dot{z} V_z -\dot{\bar{z}} V_{\bz} \right) - i
V\left(\psi\dot{\bar{\psi}}- \dot{\psi} \bp\right). \lb{onL}
\eea
This Lagrangian can be written as
\be\lb{lagrang}
{\cal L} =
\dot{Z}^A \dot{{\bar Z}}^{\bar B} g_{\bar BA} +
\left(\dot{Z}^A{\cal A}_A + \dot{{\bar Z}}^{\bar B}{\cal A}_{\bar
B}\right)\,,
\ee
where
\bea
 g_{\bz z} = {4\rho^2\over V}\left(1-\psi \bp\,{V_zV_{\bz} \over AV}\right), \  \
g_{\bp\psi} =-{4\rho^2 \over A}\left(1-\psi \bp\,{B\over A}\right), \nonumber
\eea
\bea
 g_{\bz \psi} = -4\rho^2{V_{\bz} \over AV} \bp\,, \quad
g_{\bp z} = 4\rho^2{V_{ z}\over AV} \psi\,, \lb{metric}
 \eea
while the connection terms are given by \p{connections}. It is easy to check that $g_{{\bar B} A}$
is indeed the inverse of \p{inversemetric'}.

The last topic of this Section is target space gauge transformations. The superfield Lagrangian in \p{a} is covariant,
up to a total derivative, under the following holomorphic reparametrizations compatible with the chirality of $\Phi, \Psi$:
\bea
&& \delta \Phi = \lambda(\Phi)\,, \;\;\delta \bar\Phi = \bar{\lambda}(\bar\Phi)\,, \;\;
\delta \Psi = - \frac{\partial \lambda}{\partial \Phi}\,\Psi\,, \;\;\delta \bar\Psi =
- \frac{\partial \bar\lambda}{\partial \bar\Phi}\,\bar\Psi\,, \lb{Rep1} \nn
&& \delta K = 0\,, \quad \delta V = \left(\frac{\partial \lambda}{\partial \Phi}
+  \frac{\partial \bar\lambda}{\partial \bar\Phi}\right)V\,. \lb{Rep2}
\eea
These superfield transformations induce similar ones for the component quantities:
\be
\delta z = \lambda(z)\,, \quad \delta \psi = -\lambda_z\,\psi\,, \quad \delta K(z,\bar z) = 0\,, \quad
\delta V(z, \bar z) = \left(\lambda_z + \bar\lambda_{\bar z}\right) V(z, \bar z)\,. \lb{Rep3}
\ee
It is easy to check that the on-shell Lagrangian \p{onL} is covariant under
these target reparametrizations. It is also instructive to give the transformation properties of the covariant
derivatives ${\cal P}_A$
\be
\delta {\cal P}_z = -\lambda_z{\cal P}_z + \lambda_{zz}\psi{\cal P}_\psi\,, \;
\delta {\cal P}_{\bar z} = -\bar\lambda_{\bar z}{\cal P}_{\bar z} + \bar\lambda_{\bar z \bar z}\bar\psi{\cal P}_{\bar \psi}\,, \;
\delta {\cal P}_\psi = \lambda_z {\cal P}_\psi\,, \;\delta {\cal P}_{\bar\psi} = \bar\lambda_{\bar z} {\cal P}_{\bar \psi}\,.
\ee

It is worth noting that \p{a} is also invariant under the K\"ahler-type transformations:
\be
K'(\Phi, \bar\Phi) = K(\Phi, \bar\Phi) + \Omega(\Phi) + \bar\Omega(\bar\Phi)\,.
\ee
Correspondingly, \p{onL} is shifted by a total derivative under the redefinition:
\be
K'(z, \bar z) = K(z, \bar z) + \Omega(z) + \bar\Omega(\bar z)\,. \lb{Kahl}
\ee
The supermetric \p{metric} and its inverse \p{inversemetric'}
are invariant
under these shifts, while the ``connections'' ${\cal A}_z, {\cal A}_{\bar z}$ defined in \p{connections}
undergo the target gauge transformations:
\be
\delta{\cal A}_z = i\Omega_z\,, \quad \delta{\cal A}_{\bar z} = -i \bar\Omega_{\bar z}\,. \lb{Kahlg}
\ee

\setcounter{equation}{0}
\section{${\cal N}=2$ supercharges and quantization}
 Because of the presence of arbitrary functions in the model,  we are expecting to encounter operator ordering ambiguities
 in the process of quantization.  To unambiguously determine
 the quantum Hamiltonian, we will find the supercharges of the classical system and then define
 their quantum versions in such a way that the involution $Q \rightarrow {\bar Q}$ of
 the ${\cal N} = 2$ supersymmetry algebra becomes hermitian conjugation of the quantum system.
The quantum Hamiltonian can be read off from the anticommutator of the corresponding supercharges.
In this Hamiltonian, the coefficients of the terms having the second order in the derivatives with respect to the target
space variables  should be identical with the coefficients of the terms bilinear in semi-classical momenta in \p{H}.

The  Lagrangian \p{L} transforms into a total derivative under the transformations \p{14}:
\be
\delta{\cal L}_{comp}  = {\frac{d}{dt}}\left [ -i\epsilon\left ( \chi K_z + V_z \chi \psi {\bar \psi}
+Vh{\bar \psi} + 2\rho\chi h\right) + c.c.\right].
\ee
Therefore the Noether supercharges are
\bea
Q&=&\chi\left(P_z -iK_z -i V_z \psi {\bar \psi}- 2i\rho h\right) + h\left ( P_{\psi} - i V{\bar \psi}\right), \nn
{\bar Q}&=&{\bar \chi}\left(P_{\bar z} +iK_{\bar z} +i V_{\bar z} \psi {\bar \psi}+ 2i\rho {\bar h}\right)
- {\bar h}\left ( P_{\bar \psi} - i V{ \psi}\right),\label{QQbar}
\eea
where $\chi, \bar\chi, h$ and $\bar h$ are given by the expressions \p{P}. After some work, these  supercharges can be rewritten as
\bea\label{SUSYcharges}
Q= {1\over 2i\rho}{\cal P}_z{\cal P}_{\psi}, \quad {\bar Q}= {1\over 2i\rho}{\cal P}_{\bar \psi}{\cal P}_{\bar z},\label{QQbar}
\eea
where the classical covariant derivatives ${\cal P}_A, {\cal P}_{\bar A}$ were defined in \p{CovDer}.
So the following Poisson brackets will be useful:
\be
\left\{z, P_z\right\}_{PB}= 1,\  \left\{{\bar z}, P_{\bar z}\right\}_{PB} = 1, \quad \left\{\psi, P_\psi\right\}_{PB}= -1,\
\left\{{\bar \psi}, P_{\bar \psi}\right\}_{PB} = -1\,,
\ee
Under these brackets, the covariant derivatives obey the relations:
\be
\left\{{\cal P}_A , {\cal P}_{\bar B}\right\}_{PB}= -2i \partial_A\partial_{\bar B} {\cal K}, \quad
\left\{{\cal P}_A , {\cal P}_{ B}\right\}_{PB} = 0 \label{B_fields}.
\ee
where the potential ${\cal K}$ is defined in \p{CalK}. Evaluating the brackets between the supercharges according to the usual rules
we obtain
\bea\label{QQ=H}
&\left\{Q, {\bar  Q}\right\}_{PB} = -{1\over 4 \rho^2}[{\cal P}_{ z} \left\{{\cal P}_\psi, P_{\bar \psi}\right\}_{PB} {\cal P}_{\bar z}
 -  {\cal P}_{ z}{\cal P}_{\bar \psi} \left\{{\cal P}_\psi, P_{\bar z}\right\}_{PB}\nn
 & - \left\{{\cal P}_z, P_{\bar \psi}\right\}_{PB}{\cal P}_{\bar z}{\cal P}_{ \psi}
 - {\cal P}_{ \bar \psi} \left\{{\cal P}_z, P_{\bar z}\right\}_{PB} {\cal P}_{\psi }] \nn & = -2iH_{class}\,,
\eea
where $H_{class}$ was defined in \p{H}\footnote{It is curious that the inverse supermetric $g^{A\bar B}$ entering \p{H} is related by \p{QQ=H}
through \p{B_fields} to the second derivatives of the potential ${\cal K}$:
$$
\partial_A\partial_{\bar B} {\cal K} = -4\rho^2 \epsilon_{AB}\,\epsilon_{\bar B \bar C}\,g^{B\bar C}\,,
$$
where $\epsilon_{AB}, \epsilon_{\bar A \bar B}$ are {\it symmetric} constant tensors with the only non-zero entries
$\epsilon_{z\psi} =\epsilon_{\psi z} = 1$ and $\epsilon_{\bar z \bar \psi} =\epsilon_{\bar \psi \bar z} = 1\,$, respectively. This
is an indication that the underlying geometry of our general ${\cal N}=2$ super LM is an interesting modification
of the super K\"ahler geometry, such that it is the inverse metric which is expressed through second derivatives of some scalar potential,
not the standard metric as in the (super)K\"ahler case.} and
\be
\left\{Q, {  Q}\right\}_{PB} =\left\{{\bar Q}, {\bar  Q}\right\}_{PB} = 0.
\ee

We are going to pursue the quantization of this system in the following way. We first replace
\be \label{quantization}
P_A \rightarrow - i\partial_A  , \qquad P_{\bar B} \rightarrow -i\partial_{\bar B}.
\ee
We tackle the quantization ordering  ambiguities by focusing on the definition of  $Q$ given by \p{QQbar}:
As expressed in terms of ${\cal P}_A$, the supercharge $Q$ does not exhibit any ordering ambiguities. Then we are led
to introduce a general inner product on the target superspace $(z, \bar z, \psi, \bar\psi)$ with a general measure,
and to define  $Q^\dagger$ as an Hermitian conjugate of $Q$ with respect to this inner product.

The inner product is defined as
\be\label{norm'}
< f , g > = \int \, dz\, d\bz\, d\psi\, d\bp \: F\,{\overline {\left(f \left(z,{\bar z} ,\psi, {\bar \psi}\right)\right)}}g
\left(z,{\bar z} ,\psi, {\bar \psi}\right),
\ee
where the measure $F$ is assumed to have the following $\psi, \bar\psi$ expansion:
\be
F = F_0 \left(z,{\bar z} \right) + {\bar \psi}\psi F_3 \left(z,{\bar z} \right) ,
\ee
with the real functions $F_0$ and $F_3$ to be determined. The superfunction $f$ has the generic $\psi, \bar\psi$ expansion:
\be\label{superfunction}
f\left(z,{\bar z},\psi, {\bar \psi}\right) = f_0(z, \bar z) +\psi f_1 + {\bar \psi} f_2(z, \bar z) + {\bar \psi}\psi f_3(z, \bar z)\,,
\ee
and similarly for $g\,$.

Now we wish to compute the hermitian conjugates of the basic operators with respect to this general inner product.
We note that  the anticommuting variables are always standing on the left, so to compute the component norms we will never
need to ascribe a definite Grassmann parity to the component fields. With this in mind, we derive:
\be
\left(\partial_\psi\right)^\dagger =\partial_{\bar \psi} + \psi {F_3\over F_0},\quad \left(\partial_z\right)^\dagger =-\partial_{\bar z}
-  {\partial_{\bar z}F_0\over F_0}- {\bar \psi}\psi{F_3\over F_0}\left( {\partial_{\bar z}F_3\over F_3}
-{\partial_{\bar z}F_0\over F_0}\right). \lb{modif}
\ee

The quantum version of ${\cal P}_{\bar B}$  can be obtained in a similar way, i.e. through hermitian conjugation of  ${\cal P}_{B}$
with respect to the above inner product.
As a result, the quantum version of $Q^\dagger$ will be expressed in terms of the quantum versions of ${\cal P}_{\bar B}$ according
to eq. \p{CovDer}, but with the properly modified connection terms (due to \p{modif}). This modification will change
the quantum version of eq. \p{B_fields}.
Then the quantum version of equation \p{QQ=H} will imply constraints on the measure $F$, so as to preserve
the form of ``kinetic'' terms in the quantum Hamiltonian (i.e., terms bilinear in the partial derivatives)
because the ordering procedure cannot  modify the coefficients of these highest-order terms. These coefficients
are specified by the quantum version of the relations \p{B_fields}, and  \p{QQ=H}. Requiring them
to coincide with those in the classical Hamiltonian implies the measure to be trivial,
\be
F_3 = 0\,, \quad F_0  = \omega(z)\,\bar\omega(\bar z)\,, \lb{meas1}
\ee
where $\omega(z)$ is an arbitrary
holomorphic function. In the inner product \p{norm'}, the holomorhic and antiholomorphic factors in $F_0$
can always be absorbed into the redefinition
of the superfunctions $f$ and $g$, and so, without loss of generality, we can choose $F_0 = 1$.

Having such a constant measure comes as both a bonus and a surprise. It is a bonus  because with such a measure both $Q$ and $Q^\dagger$
are naturally on the same footing. Otherwise, it would be difficult
to explain why we start by quantizing $Q$ and then define $Q^\dagger$, and not the other way around.
It is a surprise because both the Lagrangian
and the Hamiltonian in the general case involve a non-trivial supermetric.
In the quantum models associated with homogeneous superspaces as the targets, the integration measure
in the inner product can be naturally constructed with the help of the supervolume form, by requiring this measure
to be invariant under the group of super-isometries of the target (see e.g. \cite{ACIMT}). In our case the only prerequisite symmetry
of the Lagrangian and Hamiltonian is ${\cal N}=2$ supersymmetry, the transformations of which involve
the momenta (time derivatives of the coordinates). No isometry acting only on the coordinates is assumed a priori.
This invalidates the usual arguments for the construction of the invariant measure through the standard supervolume form.
Note that under the holomorphic
reparametrizations \p{Rep3} the flat measure transforms as,
\be
\delta (dz d\bar z d\psi d\bar\psi) =
2(\lambda_z + \bar\lambda_{\bar z})(dz d\bar z d\psi d\bar\psi)\,.
\ee
This can be cancelled by assuming that the (anti)holomorphic factors in \p{meas1} transform as \footnote{Alternatively,
one can take $F=1$ and choose the proper transformation laws for the superfunctions in \p{norm'} to keep the inner product invariant.}
$$
\delta \omega(z) = -2\lambda_z \omega(z)\,, \quad \delta \bar\omega(\bar z) = -2\bar\lambda_{\bar z} \bar\omega(\bar z)\,.
$$
This gives additional evidence why there is no need to insert the standard $\sqrt{{\rm sdet} g}$ factor into the definition
of the superspace measure in the case under consideration. In principle, using the ordering ambiguities, one can arrange the quantum theory in such
a way that the measure will involve a non-trivial factor (see Section 5). However, the final answers will be the same
as in the present case.

The quantum version of our  covariant derivatives ${\cal P}_A, {\cal P}_{\bar B}$ will be
\be\begin{array}{ll}
{\cal P}_z = -i(\partial_z+K_z+\psi \bar\psi V_z)\,,\quad & {\cal P}_{\bar z} = -i(\partial_{\bar z}-K_{\bar z}-\psi\bar\psi V_{\bar z})\,,\\
{\cal P}_\psi = -i(\partial_\psi + \bar\psi V) \,,& {\cal P}_{\bar \psi} = -i(\partial_{\bar \psi} + \psi V),
\end{array}
\ee
and, correspondingly, the non-vanishing relations in \p{B_fields} become
\be
[{\cal P}_z, {\cal P}_{\bar z}] = 2 \left(K_{z\bar z} +\psi\bar\psi\,V_{z\bar z}\right), \;\{{\cal P}_\psi, {\cal P}_{\bar\psi}\} = - 2V\,, \;
[{\cal P}_z, {\cal P}_{\bar\psi}] = -2 \psi\,V_z\,, \; [{\cal P}_\psi, {\cal P}_{\bar z}] = 2 \bar\psi\,V_{\bar z}\,.
\ee
Now it is straightforward to compute the quantum Hamiltonian
 \be
 H_q= {1\over 4\rho^2}[ {\cal P}_zV{\cal P}_{\bar z} + {\cal P}_z{\cal P}_{\bar \psi}V_{\bar z}{\bar \psi}
 - V_z\psi{\cal P}_{\bar z}{\cal P}_\psi +{\cal P}_{\bar \psi}(K_{z\bar z} +\psi{\bar \psi}V_{z\bar z}){\cal P}_{\psi}]\,.\lb{Hq1}
\ee

With this hermitian Hamiltonian at hand we turn to the study of the eigenvalue equation
\be
H_q\Psi\left( z,{\bar z},\psi,{\bar \psi}\right) = \lambda\Psi\left( z,{\bar z},\psi,{\bar \psi}\right),
\ee
where $\Psi$ is assumed to have general $\psi, \bar\psi$ expansion \p{superfunction},
\be
\Psi\left( z,{\bar z},\psi,{\bar \psi}\right) = f_0(z, \bar z) +\psi f_1 + {\bar \psi} f_2(z, \bar z)
+ {\bar \psi}\psi f_3(z, \bar z)\,. \lb{Psi}
\ee
An important property of this Hamiltonian is that it does not mix any  components
of $ \Psi \left( z,{\bar z},\psi,{\bar \psi}\right)$ which are  linear in $\psi, {\bar \psi}$, i.e.
\be
H_q\,\psi f_1(z, \bar z)= \lambda_1\,\psi f_1(z, \bar z)\,, \quad H_q\,\bar\psi f_2(z, \bar z)= \lambda_2\,\bar\psi f_2(z, \bar z)\,.
\ee
Or, in the component form,
\be\label{eq1'}
-{1 \over 4\rho^2} \left(\partial_{\bar z} -  K_{\bar z} \right )V\left(\partial_{z} + K_{ z}\right) f_1
=  \lambda_1\,f_1,
\ee
and
\be\label{eq2'}
-{1 \over 4\rho^2} \left(\partial_{ z} +  K_{z} \right)V
\left(\partial_{ \bar z} - K_{ \bar z} \right) f_2= \lambda_2\,f_2.
\ee
In other words, the corresponding subspaces are invariant subspaces of $H_q$.
We can also find  another pair of invariant subspaces of $H_q$
which  consist of components  of the
wave superfunction  $ \Psi \left( z,{\bar z},\psi,{\bar \psi}\right)\,$ which are even in $\psi, {\bar \psi}$. We represent $\Psi$
in \p{Psi} as a sum
\be
 \Psi = \Psi^L + \Psi^H\,, \lb{Rep1}
\ee
where
\be\label{definition'}
\Psi^L=f_0^L + {\bar \psi}\psi V f_0^L  + {\bar \psi} f_2 \equiv \Psi^L_{even} + {\bar \psi} f_2   \,,\
\Psi^H=f_0^H  - {\bar \psi}\psi V f_0^H  + \psi f_1 \equiv \Psi^H_{even} + {\psi} f_1  \,.
\ee
This corresponds just to rearranging the component fields in \p{Psi} as
\be
f_0 = f_0^L + f_0^H, \quad f_3 = V\left(f_0^L - f_0^H \right).
\ee
The superfunctions $\Psi^L_{even}$ and $\Psi^H_{even}$ also prove to be invariant subspaces under the action of $H_q$,
\be
H_q\,\Psi^L_{even} = \lambda_3\,\Psi^L_{even}\,, \quad H_q \,\Psi^H_{even} = \lambda_4\,\Psi^H_{even}\,.
\ee
This gives rise to the other two eigenvalue equations completing \p{eq1'} and \p{eq2'}:
\be\label{eq4'}
   -{1 \over 4\rho^2} \left(\partial_{ z} +  K_{z} \right)
\left(\partial_{ \bar z} - K_{ \bar z}\right)Vf_0^L= \lambda_3f_0^L
\ee
and
\be\label{eq3'}
   -{1 \over 4\rho^2} \left(\partial_{\bar z} -  K_{\bar z}\right)\left(\partial_{z} + K_{ z}  \right)Vf_0^H= \lambda_4f_0^H\,.
\ee
Thus, passing to the parametrization \p{Rep1}, \p{definition'} of the general wave superfunction  reduces
the  diagonalization of the Hamiltonian $H_q$  to two ordinary eigenvalue problems.

Indeed, by the factorization lemma which states that the non-zero eigenvalues of the operators ${\cal B} {\cal C}$ and
${\cal C} {\cal B}$ are the same (see for example  \cite{FV} \footnote{Let $H = {\cal B}{\cal C}$,
$\tilde{H} = {\cal C}{\cal B}$ and $H\psi_\lambda = \lambda \psi_\lambda, \lambda \neq 0\,$.
Then $\tilde\psi_\lambda \equiv {\cal C}\psi_\lambda$ is the eigenfunction of $\tilde{H}$ with the same eigenvalue,
$\tilde{H}\tilde\psi_\lambda
=  {\cal C}{\cal B}{\cal C}\psi_\lambda = \lambda \tilde\psi_\lambda$, i.e. $H$ and $\tilde H$ possess the same spectrum.}),
it can be easily seen
that the non-zero eigenvalues of the operators in \p{eq1'} and \p{eq4'} coincide.  The same is true for the operators in
\p{eq2'} and  \p{eq3'}.\footnote{ The corresponding pairs of operators are, respectively,
${\cal B} = \frac{1}{2i\rho}(\partial_{\bar z} - K_{\bar z})V\,, \; {\cal C} =  \frac{1}{2i\rho}(\partial_{z} + K_{z})$
and ${\cal B} = \frac{1}{2i\rho}(\partial_{z} + K_{z})V\,, \; {\cal C} =  \frac{1}{2i\rho}(\partial_{\bar z} - K_{\bar z})$.}
This is a consequence of the fact that these states are transformed into each other by the ${\cal N}=2$
supersymmetry transformations (see below).

With $F =1\,$, the inner product \p{norm'}  of the component functions,  in terms of the invariant states
of the Hamiltonian $H_q$ described
 above, is as follows:
 \bea
< f , g > &=& \int dz\,d\bz\,d\psi\,d\bar\psi \left(\bar\Psi^L\Psi^L + \bar\Psi^H\Psi^H \right) \nn
&=& \int \, {dz\, d\bz} \left({\bar f}_1g_1-{\bar f}_2g_2+2V{\bar f}_0^Lg_0^L-2V{\bar f}_0^Hg_0^H\right). \label{compnorm'}
\eea
The corresponding norm, $< f, f>\,$ is diagonal and, evidently, the norms of  states corresponding to $f_0^H$ and $f_2$
appear with the wrong sign. Therefore, like in the previous cases \cite{CIMT,ACIMT}, in order to restore the positive definiteness
we are led to introduce the metric operator
 \be
G =\cfrac{\left[{\cal P}_{\bar \psi},{\cal P}_{ \psi}\right]}{2V} +2\left( {\psi{\partial\over \partial\psi}}
- {{\bar \psi}{\partial\over\partial{\bar \psi}}}\right)  \,.\lb{metricG}
\ee
This metric operator commutes with $Q$ and $Q^\dagger$,
\be
\left[ G, Q\right] = \left[ G,{Q}^\dagger\right] = 0,
\ee
and it is a constant of motion by itself. Under the new inner product
\be
<<f, g>>= <Gf,g>, \lb{redefnorm}
\ee
the operators appearing in formulas \p{eq1'}, \p{eq2'}, \p{eq3'}, and \p{eq4'} are hermitian positive-definite operators.
It follows that their eigenvalues must be $\geqslant  0$, and the possible zero modes (specifying the ground state wave functions)
are related to solutions of the equations
\be\label{zeromodes}
\left(\partial_{z} + K_{ z}\right)g = 0, \quad\left(\partial_{ \bar z} - K_{ \bar z} \right)h = 0.
\ee
Notice that the superwave functions $\Psi^L_{even}, \; \bar\psi f_2\,$, $\Psi^H_{even}, \; \psi f_1$ corresponding
to the invariant subspaces of $H_q$
are mutually orthogonal with respect to \p{compnorm'} and \p{redefnorm}, as should be. Actually, the only effect
of passing to the new inner product is the change of the minus signs to the plus signs in the component expression
\p{compnorm'}, i.e. the change of the relative sign between terms related to each of the two irreducible ${\cal N}=2$ multiplets
(the signs between products or norms of the fields belonging to the same multiplet cannot alter  because the metric operator $G$
commutes with the ${\cal N}=2$ supersymmetry generators).

It is also worthwhile to note that the eigenvalue equations \p{eq1'}, \p{eq2'}, \p{eq4'}, and \p{eq3'}
are covariant under the holomorphic reparametrizations \p{Rep3}, and the  K\"ahler-type transformations \p{Kahl},
if the wave functions are assumed
to transform as
\bea
&&\delta f_0^L = -\left(\lambda_z + \bar\lambda_{\bar z} + \Omega - \bar\Omega\right)f^L_0\,, \quad
\delta f_0^H = -\left(\lambda_z + \bar\lambda_{\bar z} + \Omega - \bar\Omega\right)f^H_0\,, \nn
&& \delta f_1 = -\left(\bar\lambda_{\bar z}+ \Omega - \bar\Omega\right)f_1\,, \quad \delta f_2 =
-\left(\lambda_{z}+ \Omega - \bar\Omega\right)f_2\,.
\eea
It is straightforward to check that the inner products \p{compnorm'} and \p{redefnorm} are invariant under these target gauge transformations.

As the last topic of this Section, we shall study the action of the supersymmetry generators on the invariant subspaces
of the Hamiltonian which we described above.

 We have:
 \bea\label{q1}
Q\Psi^L = 0, &{Q}^\dagger\Psi^L ={i\over 2\rho}
[({\partial_{\bar z}} - K_{\bar z})f_2 + 2 \psi ({\partial_{\bar z}}
- K_{\bar z})Vf_0^L \nn & -{\bar \psi}\psi V({\partial_{\bar z}} - K_{\bar z})f_2], \lb{I}
 \eea
 and
  \bea\label{q2}
{Q}^\dagger\Psi^H = 0, &{ Q}\Psi^H =
{i\over 2\rho} [({\partial_{ z}} +K_{ z} )f_1 + 2{\bar \psi}
({\partial_{z}} + K_{ z} )Vf_0^H \nn & +{\bar \psi}\psi V({\partial_{ z}} +K_{ z})f_1]. \lb{II}
 \eea
Now it is easy to see that the general superfunction $\Psi$ contains two irreducible
${\cal N} =2$ multiplets $(f_1, f_0^L)$ and $(f_2, f_0^H)\,$,
which, before the redefinition of the norm, have positive and negative norms, respectively.
Defining the ${\cal N} =2$ supersymmetry transformation
of the general wave function $\Psi = \Psi^L + \Psi^H$ as
\be
\delta \Psi = (\epsilon Q + \bar\epsilon Q^\dagger) \Psi\,,
\ee
we find from \p{I} and  \p{II}
\bea
&&  \delta f_0^L = \frac{i}{2\rho}\, \epsilon\, ({\partial_{ z}} +K_{ z} )\,f_1\,, \quad \delta f_1 =
-\frac{i}{\rho}\,\bar\epsilon\, ({\partial_{\bar z}}- K_{\bar z})V\, f_0^L\,, \nn
&& \delta f_0^H = \frac{i}{2\rho}\, \bar\epsilon\, ({\partial_{\bar z}} - K_{ \bar z} )\,f_2\,, \quad \delta f_2 =
-\frac{i}{\rho}\,\epsilon\, ({\partial_{z}} + K_{ z})V\, f_0^H\,. \lb{N2susyWF}
\eea

The ground state wave superfunctions $\Psi^L_{vac}, \Psi^H_{vac}$ are defined as zero eigenvalues
 of $H_q$. The corresponding wave functions are solutons of eqs. \p{zeromodes}, so  $\Psi^L_{vac}, \Psi^H_{vac}$ automatically obey
the conditions
\be
Q \Psi^L_{vac} = Q^\dagger \Psi^L_{vac} = Q \Psi^H_{vac}
= Q^\dagger \Psi^H_{vac} = 0\,, \lb{N2susyVAC}
\ee
as a consequence of the relations \p{I}, \p{II}. The set of ground states is spanned by two holomorphic and two antiholomorphic functions:
\bea
&& (f_2)_{vac} = e^K \tilde{f}_2(z)\,, \; (f^L_0)_{vac} = V^{-1} e^K \tilde{f}^L_0(z)\,, \nn
&& (f_1)_{vac} = e^{-K} \tilde{f}_1(\bar z)\,, \; (f^H_0)_{vac} = V^{-1} e^{-K} \tilde{f}^H_0(\bar z)\,. \lb{vac}
\eea
Using the transformation properties \p{N2susyWF}, it is straightforward to check that the functions \p{vac} are
indeed singlets under ${\cal N}=2$ supersymmetry.

Finally, we notice that the obvious requirement of  finiteness for the $z, \bar z$ integrals present in the definition
of the inner products \p{compnorm'} and \p{redefnorm}, and of the corresponding norms, imposes rather severe
restrictions on the asymptotic behavior of the
admissible class of wave functions $f_0^L, f^H_0, f_1$ and $f_2$ as $z, \bar z \rightarrow \infty$, as well as on the admissible choice
of the potentials $K(z, \bar z)$ and $V(z, \bar z)$. This issue is difficult to analyze in general.
We shall discuss it on a concrete example in Sect. 7.

\setcounter{equation}{0}
\section{Alternative quantization}\label{second_case}
As follows from the above analysis, demanding the identity of the coefficients of the terms quadratic in the momenta
in the quantum and classical versions  of the Hamiltonian imposes  very stringent conditions
on the hermitan adjoint properties of the covariant derivatives. Namely, $ {\cal P}_A^\dagger $ computed within the natural inner product
should give rise to $ {\cal P}_{\bar A}$, which in turn forces the integration measure in the inner product
to be almost constant (see \p{meas1}). It is interesting that this chain of requirements can be relaxed by considering
an equivalent classical form
of the supersymmetry charges $Q$ and $\bar Q$. Indeed, consider the equivalent classical expression for the supersymmetry generators:
\bea\label{SUSYcharges'}
Q'= {1\over 2i\rho}{\cal P'}_z{\cal P}_{\psi}, \quad {\bar Q}'= {1\over 2i\rho}{\cal P}_{\bar \psi}{\cal P'}_{\bar z},\label{QQbar'}
\eea
where
\be\begin{array}{ll}
{\cal P'}_z = (P_z -iK_z - {V_z\over V}\psi P_{\psi})\,,\quad & {\cal P'}_{\bar z} = (P_{\bar z}
+ iK_{\bar z} -{V_{\bar z}\over V}{\bar \psi} P_{\bar \psi}),\\
{\cal P}_\psi = (P_\psi -i \bar\psi V) \,,& {\cal P}_{\bar \psi} = (P_{\bar \psi} - i\psi V).
\end{array}
\ee
It is easy to see that ${\cal P}_z - {\cal P}'_z  \sim \psi{\cal P}_\psi$, and therefore the classical supercharge $Q$ is not modified, and
a similar argument is valid for $\bar Q$. The corresponding classical brackets among these ``new covariant derivatives'' can be easily
obtained from \p{B_fields}.

In this Section we are going to use the  quantization rules  (\ref{quantization}) and $1\over V$ as a measure in (\ref{norm'}).
We are also going
to use the following quantum ordering prescription in the definitions:
\be\begin{array}{ll}
{\cal P'}_z = -i(\partial_z + K_z - {V_z\over V}\psi \partial_{\psi} + a {V_z\over V})\,,\quad & {\cal P'}_{\bar z}
= -i(\partial_{\bar z}  - K_{\bar z} -{V_{\bar z}\over V}{\bar \psi} \partial_{\bar \psi} - a{V_{\bar z}\over V}),\\
{\cal P}_\psi = -i(\partial_\psi + \bar\psi V) \,,& {\cal P}_{\bar \psi} = -i(\partial_{\bar \psi} + \psi V),
\end{array}
\ee
where the extra terms with a real constant  $a$ in the expression of the $\cal P'$s reflects the ordering ambiguity
in the products $\psi
P_{\psi}$ and ${\bar \psi}{\bar P}_{\bar \psi}$.

The following statements can be checked to be true
\be
{\cal P'}_z^{\dagger} = {\cal P'}_{\bar z}, \quad  {\cal P}_\psi^{\dagger}= - {\cal P}_{\bar \psi}.
\ee
Therefore, if  the supercharge $Q'$  is ordered as in (\ref{SUSYcharges'}), and $Q'{}^\dagger$ is defined as the hermitian conjugate
of $Q'$ with respect to the inner product with the measure $1/V$,
we can again implement the involution of the abstract ${\cal N}=2$ superalgebra as the hermitian conjugation of the quantum operators.
It remains to check that the coefficients of the  terms quadratic in the momenta of  the quantum and classical Hamiltonians are equal.
The algebra of the new covariant derivatives is
\bea
& \{ {\cal P}_\psi\, , {\cal P}_{\bar \psi} \}= -2 V \,,\ \
[{\cal P'}_z\, , {\cal P'}_{\bar z} ]= 2K_{z\bar z} + \left( \partial_z\partial_{\bar z} \ln V\right)
( \bp \partial_\bp -  \psi \partial_{ \psi} +2a ), \nn
&[ {\cal P'}_z\, , {\cal P}_{\bar \psi} ]=0,\quad [{\cal P'}_{\bar z}\,, {\cal P}_{ \psi}]=0,\quad
{\cal P}_\psi {\cal P}_{ \psi}= {\cal P}_{\bar \psi} {\cal P}_{ \bar \psi}=0,    \nn
&[  {\cal P'}_z\, , {\cal P}_{ \psi}]= - i\left( \partial_z \ln V\right){\cal P}_{ \psi} \,,\quad [ {\cal P'}_{\bar z}\, ,
{\cal P}_{\bar \psi}] = -i\left(\partial_{\bar z} \ln V\right) {\cal P}_{\bar \psi}.\label{primecov}
\eea
The quantum Hamiltonian reads
 \bea\label{Hq}
2 \tilde{H}_q=\left\{Q', Q'{}^{\dagger}\right\} = {1\over 2\rho^2}\left[{\cal P'}_z V{\cal P'}_{\bar z}
+{\cal P}_{\bar \psi}\left( K_{z{\bar z}} + {1\over 2}{\partial}_z{\partial}_{\bar z}\ln V\left( {\bar \psi}{\partial_{\bar \psi}}
- { \psi}{\partial_{ \psi}} +2a\right)\right){\cal P}_{ \psi}\right].
\eea
An inspection of this expression reveals that it contains terms which formally appear as having  three odd derivatives.
Upon rewriting them in detail,
because of the ordering, these terms generate an additional term in the product of the two odd momenta. By choosing $a = - {1 \over 2}$
one can cancel the additional contribution to ensure that the coefficients
of the momenta-squared terms in the quantum Hamiltonian  are identical to those in the classical Hamiltonian.
Should we have chosen another ordering  instead of the one in \p{SUSYcharges'}, for example the symmetrical (Weyl) prescription
${1\over 2}\left({\cal P'}_z{\cal P}_{\psi} +{\cal P'}_{\psi}{\cal P}_{z}\right)$ with the corresponding definition
of  ${Q'}^\dagger$, it can be shown (with the help of \p{primecov}) to require a different  value of $a$.
In what follows we are  going  to pursue the consequences of the ordering chosen in  \p{SUSYcharges'}.

Now, proceeding as we did  in the previous Section, we obtain the same invariant subspaces of the new quantum Hamiltonian.
Using the same expansion for the relevant wave superfunction, we derive
\be\label{eq1}
\tilde{H}\psi f_1 =  -{1 \over 4\rho^2} V\left(\partial_{\bar z}
-  K_{\bar z} + {1\over 2}\partial_{\bar z}\ln V\right)\left(\partial_{z}
+ K_{ z} -{1\over 2}\partial_{ z}\ln V\right){ \psi}f_1= \lambda_1{ \psi}f_1,
\ee
and
\be\label{eq2}
\tilde{H}{\bar \psi}f_2 =  -{1 \over 4\rho^2} V\left(\partial_{ z} +  K_{z} + {1\over 2}\partial_{ z} \ln V\right)
\left(\partial_{ \bar z} - K_{ \bar z} -{1\over 2} \partial_{\bar z} \ln V\right){ \bar \psi}f_2= \lambda_2{ \bar \psi}f_2.
\ee
Then, using \p{definition'}, we obtain the other set of invariant subspaces
\be\label{eq3}
\tilde{H}f_0^H =   -{1 \over 4\rho^2} \left(\partial_{\bar z} -  K_{\bar z}
   - {1\over 2}\partial_{\bar z}\ln V\right)V\left(\partial_{z} + K_{ z} +{1\over 2}\partial_{ z}\ln V\right)f_0^H= \lambda_3f_0^H,
\ee
and
\be\label{eq4}
\tilde{H}f_0^L =   -{1 \over 4\rho^2} \left(\partial_{ z} +  K_{z} - {1\over 2}\partial_{ z} \ln V\right)V
\left(\partial_{ \bar z} - K_{ \bar z} +{1\over 2} \partial_{\bar z} \ln V\right)f_0^L= \lambda_4f_0^L.
\ee
 The inner product \p{norm'}  of the component functions, with the measure $V^{-1}$,
 in terms of the invariant states of the Hamiltonian $H_q$ described above, is given by the integral
 \be\label{compnorm}
< f , g > = \int \, {dz\, d\bz \over V} \left({\bar f}_1g_1-{\bar f}_2g_2+2V{\bar f}_0^Lg_0^L-2V{\bar f}_0^Hg_0^H\right).
\ee
At this stage it is easy to see that, changing the functions in \p{compnorm} by
\be
f_0^L\rightarrow V^{1\over 2}f_0^L\,, \quad f_0^H\rightarrow V^{1\over 2}f_0^H\,, \quad f_i\rightarrow V^{1\over 2}f_i\,, \ (i = 1,2),
\ee
we come back to \p{compnorm'}, while the equations (\ref{eq1}) - (\ref{eq4}) are converted into
the previous set (\ref{eq1'}), (\ref{eq2'}), (\ref{eq3'}), (\ref{eq4'}). The supercharges of the different quantization
schemes are connected by the relation
\be
{V}^{-\frac{1}{2}} Q'\, V^{\frac{1}{2}} = Q, \quad {V}^{-\frac{1}{2}} Q'{}^\dagger\, {V}^{\frac{1}{2}} = Q^\dagger. \lb{simQ}
\ee
It
is also easy to find the explicit relation between the Hamiltonians $H_q$ and $\tilde{H}_q$ defined by eqs. \p{Hq1} and \p{Hq}:
\be
\tilde{H}_q = H_q + \frac{1}{8\rho^2}\left[V_{z\bar z} + \frac{1}{2}\frac{V_z V_{\bar z}}{V} + i\left(V_z{\cal P}_{\bar z}
+ V_{\bar z}{\cal P}_z \right) -i\frac{V_z V_{\bar z}}{V}\left(\psi {\cal P}_\psi +  \bar\psi {\cal P}_{\bar\psi}\right)\right].
\ee
This relation can be rewritten as the following simple similarity transformation,
\be
{V}^{-\frac{1}{2}}\tilde{H}_q\, {V}^{\frac{1}{2}} = H_q\,, \lb{simH}
\ee
which agrees with \p{simQ} and proves the equivalence of the two quantization schemes.\footnote{A similar equivalence transformation
between various quantization schemes in the conventional supersymmetric quantum mechanics and its relation to different definitions
of the inner product were discussed many years ago in \cite{ASm}.}

Now, using \p{definition'},  we have:
 \bea
Q\Psi^L = 0, &{Q}^\dagger\Psi^L ={i\over 2\rho} [({\partial_{\bar z}} - K_{\bar z}
- {1\over 2}\partial_{\bar z}\ln V)f_2 + 2 \psi V({\partial_{\bar z}} - K_{\bar z}
+{1\over 2}\partial_{\bar z}\ln V)f_0^L \nn & -{\bar \psi}\psi V({\partial_{\bar z}}
- K_{\bar z} -{1\over 2}\partial_{\bar z}\ln V)f_2], \lb{QL}
 \eea
 and
  \bea
{Q}^\dagger\Psi^H = 0, &{ Q}\Psi^H ={i\over 2\rho} [({\partial_{ z}} +K_{ z}
- {1\over 2}\partial_{ z}\ln V)f_1 + 2{\bar \psi}V ({\partial_{z}} + K_{ z}
+{1\over 2}\partial_{z}\ln V)f_0^H \nn & +{\bar \psi}\psi V({\partial_{ z}} +K_{ z} -{1\over 2}\partial_{ z}\ln V)f_1].\lb{QH}
 \eea
As follows from \p{eq1} - \p{eq4}, the ground state wave functions corresponding to zero eigenvalues of $\tilde{H}$ are defined
by the equations
\bea
&& ({\partial_{z}} + K_{z}- \frac{1}{2}\partial_{z}\ln V)\,(f_1)_{vac} =
({\partial_{\bar z}} - K_{\bar z} + {1\over 2}\partial_{\bar z}\ln V)\,(f_0^L)_{vac} = 0\,, \nn
&& ({\partial_{\bar z}} - K_{\bar z}
- {1\over 2}\partial_{\bar z}\ln V)\,(f_2)_{vac} = ({\partial_{z}} + K_{ z} +{1\over 2}\partial_{z}\ln V)\,(f_0^H)_{vac} = 0\,, \lb{grstCOND}
\eea
which imply that the ground state wave superfunctions $\Psi^L_{vac}, \Psi^H_{vac}$ are singlets of ${\cal N}=2$
supersymmetry, like in the first quantization scheme (eqs. \p{N2susyVAC}).

Finally, we note that the passing to the positive-definite inner product from \p{compnorm} in this quantization scheme
is accomplished by the same operator $G$ as in \p{metricG}, but now it should be transformed on the pattern of \p{simQ} and \p{simH}.

\setcounter{equation}{0}
\section{Factorized Schr\"odinger operators}\label{second_case}
In this Section and the next, we shall deal with the quantization scheme of Section 5, as it makes manifest
some important properties of the system under consideration.

As follows from \p{eq1} - \p{eq4},  the Schr\"odinger operator factorizes on the corresponding  invariant subspaces.
Moreover the relations \p{eq1} and \p{eq2} tell us in this case we deal with an important class
of factorizable Hamiltonians, with so-called $\beta$-factorization and $\alpha$-factorization respectively, in the terminology of
\cite{FV}. These factorizations correspond to general Schr\"odinger operators on the manifold $M_2$ with the metric
$g_{z\bar z} = V^{-1}\,, \;g^{z\bar z} = V$, and with a potential related to the corresponding magnetic fields. For simplicity, we set $4\rho^2 = 1$,
and rewrite eqs. \p{eq1} and \p{eq2} as

\begin{equation}
\;\;-g^{z\bar z}\,\bar\nabla{}^{(-)}_{\bar z}\nabla{}^{(-)}_z \,f_1 = \lambda_1\,f_1\,, \quad  \;\;
-g^{z\bar z}\,\nabla^{(+)}_z\bar{\nabla}^{(+)}_{\bar z} \,f_2 = \lambda_2\,f_2\,, \lb{5758}\tag{6.1a,b}
\end{equation}%
\begin{subequations}
\end{subequations}
where
\be
\nabla^{(\pm)}_{ z}= \partial_{z} + K_{ z} \pm {1\over 2}\partial_{ z}\ln V, \quad {\rm } \quad {\bar \nabla}^{(\pm)}_{\bar z}
 = \partial_{\bar z} -  K_{\bar z} \mp {1\over 2}\partial_{\bar z}\ln V. \lb{nablas-pm}
\ee
The eigenvalue problems \p{5758} can then be rewritten in the manifestly factorized form
\begin{subequations}
\end{subequations}
\begin{equation}
 \;\;-\Big[V^{1\over 2}\,\bar\nabla{}^{(-)}_{\bar z} - \frac{1}{2}\partial_{\bar z}
\ln V\Big]\Big[V^{1\over 2}\,\nabla{}^{(-)}_z\Big] \,f_1 = \lambda_1\,f_1\,,
\lb{5758'}\tag{6.3a}
\end{equation}
\begin{equation}
 \;\; - \Big[V^{1\over 2}\,\nabla^{(+)}_z - \frac{1}{2}\partial_{z}\ln V\Big] \Big[V^{1\over 2}\,\bar\nabla^{(+)}_{\bar z}\Big] \,f_2
= \lambda_2\,f_2\,. \tag{6.3b}
\end{equation}

A similar factorization property can be shown for the eigenvalue problems \p{eq3} and \p{eq4} (see below). So the world-line
${\cal N}=2$ supersymmetry implies the factorization property for the component Hamiltonians (modulo constant shifts and addition
of the explicit potential terms, see footnote on p.4).

A sufficient condition for systems with factorized Hamiltonians to be integrable is the existence of an infinite sequence
of factorization chains, which corresponds to determining infinite sequences of eigenvalues and eigenvectors of the corresponding
Hamiltonians. As was proved in \cite{FV}, in the case of systems on $M_2$ this condition is fulfilled if and only if {\bf i)} The
Gauss curvature $ \mathbb{K}$ of $M_2$ is a constant:
\begin{equation}
\mathbb{K} = 2g^{z \bar z}\partial_z\partial_{\bar z}\ln g^{z \bar z} = const, \lb{const1}
\end{equation}
and {\bf ii)} The corresponding magnetic field is also a constant:
\be
g^{z \bar z} \left[ \bar\nabla_{\bar z}, \nabla_{ z} \right] = c = const\,. \lb{const2}
\ee

The constancy of the Gaussian curvature $ \mathbb{K}$ on $M_2$ always implies that
the metric $ g_{{\bar z} z}$ is K\"ahler (see e.g. \cite{DNF})
\be
V^{-1} = g_{{\bar z} z} = \partial_z\partial_{\bar z}\Phi\,,\qquad
\Phi = {2 \over  \mathbb{K}} \ln g^{z \bar z} = {2 \over  \mathbb{K}} \ln V, \lb{Vcond1}
\ee
where $ \Phi$ is the K\"ahler potential and we assumed that $\mathbb{K} \neq 0\,$. The constancy condition \p{const2}
for the magnetic field in \p{eq1} (or in the equivalent
forms of this relation (\ref{5758}), (\ref{5758'},b)), with $\bar\nabla_{\bar z}^{(-)}, \nabla_{ z}^{(-)}$ from \p{nablas-pm},
requires that
\be
K_{z{\bar z}} = \frac{1}{4}\left(\mathbb{K} + c\right)\partial_z\partial_{\bar z}\Phi\,.
\ee
This equation is solved by
\be
K  = \frac{1}{4}\left(\mathbb{K} + c\right)\Phi =  \frac{1}{2}\left(1 + \frac{c}{\mathbb{K}}\right)\ln V\,, \lb{Kcond1}
\ee
up to a K\"ahler gauge transformation.

Therefore, the connection terms in eq. \p{eq1} are K\"ahler connection terms:
\be
\nabla^{(-)}_{ z}= \partial_{z} + {c\over 2 \mathbb{K}}\partial_{ z}\ln V, \quad  \quad \bar\nabla^{(-)}_{\bar z}
 = \partial_{\bar z} - {c\over 2 \mathbb{K}}\partial_{\bar z}\ln V\, . \lb{nabla-}
\ee
This also automatically applies  to the second equation \p{eq2}, where now
\be
\nabla^{(+)}_{ z}= \partial_{z} + \left(1 + {c\over 2 \mathbb{K}}\right)\partial_{ z}\ln V, \quad \quad \bar\nabla^{(+)}_{\bar z}
 = \partial_{\bar z} -  \left(1 +{c\over 2 \mathbb{K}}\right)\partial_{\bar z}\ln V\, . \lb{nabla+}
\ee
The eigenvalue problem \p{eq2} can then be cast in the $\beta$-factorized form similar to \p{eq1}, up to a constant shift:
\be
-g^{z\bar z}\,\nabla^{(+)}_z\bar\nabla^{(+)}_{\bar z} \,f_2 =\Big[-g^{z\bar z}\,\bar\nabla^{(+)}_{\bar z}\nabla^{(+)}_{z} +
\left(\mathbb{K} + \frac{c}{2}\right)\Big] \,f_2 = \lambda_2\,f_2\,.
\ee

In addition the eigenvalue problems \p{eq3} and \p{eq4} can be rewritten in a  form similar to \p{eq1}, \p{eq2}, only now with
\bea
\nabla^{(\pm)}_{ z}= \partial_{z} + \frac{1}{2} \left(1 +  {c\over \mathbb{K}}\right)\partial_{ z}\ln V\,, \quad
\bar\nabla^{(\pm)}_{\bar z} = \partial_{\bar z} - \frac{1}{2} \left(1 +  {c\over \mathbb{K}}\right)\partial_{\bar z}\ln V.\,
\eea
This follows by moving $V$ in \p{eq3}, \p{eq4} to the left and making the equivalence transformation of the wave functions as
$f_0^{H, L} = V^{-{1\over 2}} \tilde{f}_0^{H,L}\,$.\footnote{Actually,
this two-step procedure brings \p{eq3} and \p{eq2} to the $\beta$ and $\alpha$ factorized form
{\it before} imposing the infinite factorization chain conditions \p{const1}, \p{const2}.}

Thus eventually we deal with the coupling of states of different charge to a magnetic field, with some
constant shifts in the corresponding Hamiltonians.
As was already mentioned, to generate infinite sequences of eigenvalues and eigenvectors through the factorization method
the manifold $M_2$ must have a constant Gauss curvature $\mathbb{K} \,$, and the vector potentials describing the coupling
to the magnetic field must be K\"ahler connections.  Both conditions are explicitly satisfied under
the choice \p{Vcond1} and \p{Kcond1}.

It is well known (see e.g. \cite{DNF}) that, by a proper choice of coordinates,
one can bring (locally) the metric on $M_2$ to a standard form which exhibits the isometries
of the corresponding manifold. For $\mathbb{K}  > 0$ one has the metric
of the sphere $g^{z \bar z} \sim (1 + z{\bar z})^2$, and   for $\mathbb{K}  < 0$ one has the metric
of the hyperboloid $g^{z \bar z} \sim (1 - z{\bar z})^2$. For $\mathbb{K}  =0$ the metric is constant and this case corresponds
to the super planar model considered in Sect. 2. Note that the latter case requires a separate consideration, because it is degenerate
and does not directly match the $\mathbb{K} \neq 0$ analysis given above. In particular, no relation between the functions $V = 1$ and
$K(z,\bar z) = z\bar z$ (where the numerical coefficients are chosen so as to ensure correspondence with Sect. 2) arises in this case.
Eq. \p{const1} with $\mathbb{K}  =0$ is satisfied trivially, while \p{const2} is satisfied with $c=2\,$. The K\"ahler
form of the connections is built-in from the very beginning, $\nabla^{(\pm)}_z = \partial_z + \bar z\,$ and
$\bar\nabla^{(\pm)}_{\bar z} = \partial_{\bar z} - z$ in \p{nablas-pm} (with $4\rho^2 = 1$).

Finally, recall that the existence of infinite factorization chains is only
a sufficient condition for the complete integrability of the factorizable Hamiltonians.
However, because of insufficient information about   other possible integrable models,
we restrict ourselves to this option.

In the next Section  we shall elaborate on the particular case of the 2-sphere $\sim \mathbb{CP}^1$,
with the functions $ K(z, {\bar z})$ and $V(z, {\bar z})$ being chosen as
\be\label{factconditions}
K(z, {\bar z}) = - N \ln ( 1 + z{\bar z}) , \qquad V(z, {\bar z}) = g^{z \bar z} = (1 + z{\bar z})^2.
\ee
The value of  $N$ is quantized by the standard cohomology arguments, $N\in (\mathbb{N}, \mathbb{N} + {1\over 2})$, and we take the minus sign
in order to deal with the analytic sector. Also, for simplicity, we shall again set $ 4\rho^2 =1\,$.
The choice \p{factconditions} implies
\be
{c \over \mathbb{K}} = -(N+1)
\ee
in \p{Kcond1} and subsequent formulas.

\setcounter{equation}{0}
\section{ $\mathbb{CP}{}^1$ model.}
Here we consider the $SU(2)$ invariant subclass of
the actions \p{a}, with the following potentials:
\be\label{cphere_potentials}
K(\Phi, \bar\Phi) = - N\ln \left(1 + \Phi \bar\Phi\right), \quad
V(\Phi, \bar\Phi) = \left(1 + \Phi \bar\Phi\,\right)^2.
\ee
 It is easy to check that under this choice \p{a} is invariant with respect to the standard
$\mathbb{C}\mathbb{P}{}^1$ realization of the $SU(2)$ transformations:
\be
\delta \Phi = \varepsilon + i\beta\, \Phi + \bar\varepsilon\, \Phi^2\,, \quad \delta \Psi =
-\left(i\beta + 2\bar\varepsilon\,\Phi\right)\Psi\,.
\ee
Thus the superfields $\Phi$ and $\bar\Phi$ can be interpreted as the complex coordinates of $\mathbb{C}\mathbb{P}{}^1 \sim SU(2)/U(1)$,
with $K(\Phi, \bar\Phi)$ being related to the K\"ahler potential. Since the corresponding bosonic functions $K(z,\bz)$ and $V(z,\bz)$
are just of the form \p{factconditions},  we deal with the dynamics of a particle  on the sphere in a magnetic
field  --- the field of a Dirac monopole located at the center.
For this particular case the on-shell Lagrangian \p{onL} (up to a renormalization factor) reads
\bea
{\cal L}_{su(2)} &=& \frac{\dot z\dot{\bar{z}}}{\left(1 + z \bz\right)^2} +
N^{-1}\left(1 + z\bz \right)^2\left[1 +2 N^{-1}\psi\bp\left(1 + z\bz \right)^2\right] \nabla\psi \nabla\bp \nn
&& -\;i\left[\frac{N - 2\psi\bp\left(1 + z\bz\right)^2}{1 + z\bz}
\left(\dot{z} \bz - \dot{\bar{z}}z\right)
- \left(1 + z\bz\right)^2\left(\dot\psi\bp - \psi\dot{\bar{\psi}}\right)\right], \lb{onLsu21}
\eea
where
\be
\nabla\psi = \dot{\psi} + 2\frac{\dot{z} \bz}{1 + z\bz}\,\psi\,, \quad
\nabla\bp = \dot{\bar{\psi}} + 2\frac{\dot{\bar{z}} z}{1 + z\bz}\,\bp\,.
\ee
This Lagrangian can be rewritten as
\be
{\cal L}_{su(2)} = \dot{Z}^A \dot{{\bar Z}}^{\bar B} g_{\bar BA} + \left(\dot{Z}^B{\cal A}_B
+ \dot{{\bar Z}}^{\bar B}{\cal A}_{\bar B}\right),
\ee
with
\bea
&& g_{\bz z} = \frac{1}{\left(1 + z \bz\right)^2} +{ 4z\bz \over N}\, \psi\bp\,, \quad
g_{\bp\psi} ={\left(1 + z \bz\right)^2 \over N}\left[1 + {2\psi\bp\left(1 + z \bz\right)^2 \over N}\right], \nn
&& g_{\bz \psi} = {2\left(1 + z \bz\right)\over N} z\bp\,, \quad
g_{\bp z} = -{2\left(1 + z \bz\right)\over N} \bz \psi\,, \lb{g1} \\
&& {\cal A}_z = i{\bar z}\frac{-N + 2\psi\bp\left(1 + z\bz\right)^2}{1 + z\bz}\,,
\;{\cal A}_{\bar z} = i{z}\frac{N - 2\psi\bp\left(1 + z\bz\right)^2}{1 + z\bz}\,,\\
&&{\cal A}_{\psi} = i \left(1 + z\bz\right)^2{\bar \psi}, \;
{\cal A}_{\bp} = i\psi \left(1 + z\bz\right)^2.
\eea
The entries of the inverse target space metric are given by
\bea
&& g^{z \bz} = \left(1 + z \bz\right)^2, \quad
g^{\psi \bp} = \frac{1}{\left(1+ z \bz\right)^2}\left[N -
2\psi\bp\left(1 + z \bz\right)^2\left(1 + 2z \bz\right) \right], \nn
&& g^{\psi \bz} = 2\left(1 + z \bz\right) \bz \psi\,, \quad
g^{z \bp} = -2\left(1 + z \bz\right) z \bp\,.
\eea

The action corresponding to the Lagrangian \p{onLsu21} is invariant under the ${\cal N}=2$ supersymmetry transformations \p{14}, with
the auxiliary fields $h$ and $\chi$ being expressed by the general formulas \p{aux2}, and under $SU(2)$ transformations
\be
\delta z = \varepsilon + i\beta\, z + \bar\varepsilon\, z^2\,, \quad \delta \psi =
-\left(i\beta + 2\bar\varepsilon\,z\right)\psi\,. \lb{su2comp}
\ee
These invariances are the only symmetries of the considered model.

The Lagrangian \p{onLsu21} presents an ${\cal N}=2$ supersymmetric extension of the $SU(2)$ invariant bosonic Lagrangian
describing a particle in the background of a Dirac monopole placed at the center of the 2-sphere
$S^2 \sim \mathbb{CP}^1$ (and so underlying a LM on the 2-sphere $S^2$ \cite{Hald}). A new feature of this extension, as compared
with the minimal ${\cal N}=2$ extensions discussed, e.g. in \cite{D'Hoker:1983xu} - \cite{Hong:2005ce},
is that it involves the non-standard second-order
kinetic term for fermions (along with the canonical first-order term) and goes into the Lagrangian of the superplane model
in the flat limit.

Actually, like in the bosonic case, we deal with a bunch of models parametrized by the parameter $N$.
The quantization of these models follows the general pattern, and we will
specialize the general results obtained in the preceding Sections. Working within the alternative
quantization scheme, which allows a more direct comparison with the general Schr\"{o}dinger operator on a
two-dimensional manifold, the corresponding eigenvalue equations are:
\bea\label{eq1''}
 & - V \nabla_{\bar z}^{(N+1)}\nabla_{ z}^{(N+1)}f_1= \lambda_1f_1,\ \
 - V \nabla_{ z}^{(N-1)}\nabla_{ \bar z}^{(N-1)}f_2= \lambda_2f_2,\nn
  & -  \nabla_{\bar z}^{(N-1)}V\nabla_{ z}^{(N-1)}f_0^H= \lambda_3 f_0^H,\ \
  -  \nabla_{ z}^{(N+1)}V\nabla_{ \bar z}^{(N+1)} f_0^L=  \lambda_4 f_0^L,
\eea
where
\be\label{eq2''}
 \nabla_z^{(N)} = \partial_z - N {{\bar z} \over {1 + {\bar z}z}}, \ \  \nabla_{\bar z}^{(N)}
 = \partial_{\bar z} + N {{ z} \over {1 + {\bar z}z}}.
\ee

One more advantage of the alternative quantization scheme in the present case is  that  the integration measure
in the inner product \p{compnorm} is
just the $SU(2)$ invariant integration measure over $\mathbb{CP}^1$, $dz\,d\bar z / (1 + z\bar z)^2$, so requiring
the relevant wave functions to be normalizable actually amounts to the standard demand of their square-integrability
on $\mathbb{CP}^1 \sim S^2\,$, under which the function proves to be globally defined on $S^2$.
In turn, this implies that the normalizable wave functions should encompass irreducible
unitary representations of $SU(2)\,$. It is useful to know the $SU(2)/U(1)$ transformations of the wave functions
$f_1, f_2, f_0^L, f_0^H$ which leave invariant the inner product \p{compnorm} in the model under consideration:
\bea
&& \delta f_{1} = -[(N+1)(\varepsilon \bar z - \bar\varepsilon z) + \delta z \partial_z + \delta \bar z \partial_{\bar z}]\, f_1\,, \nn
&& \delta f_{2} = -[(N-1)(\varepsilon \bar z - \bar\varepsilon z) +\delta z \partial_z + \delta \bar z \partial_{\bar z}]\,f_1\,, \nn
&& \delta f_0^{L, H} = -\Big[ \varepsilon \bar z + \bar\varepsilon z + N(\varepsilon \bar z - \bar\varepsilon z) +\delta z \partial_z
+ \delta \bar z \partial_{\bar z}\Big]f_0^{L, H}\,. \lb{su2wf}
\eea

Now we shall analyze the structure of the wave functions as solutions of \p{eq1''} - \p{eq2''}. It turns out that this structure
essentially depends on the value of $N \in (\mathbb{N}, \mathbb{N} + {1\over 2})$. The normalizability requirement imposes rather
severe restrictions on the admissible choice of the wave functions.
\vspace{0.2cm}

\noindent{\bf Ground states}
\vspace{0.2cm}

We start our analysis with the ground states. From the point of view of the underlying bosonic Landau model on $S^2 \sim \mathbb{CP}^1$ \cite{Hald},
they correspond to the lowest Landau level (LLL). The LLL wave functions are defined by the equations \p{grstCOND} specialized
to the case under consideration:
\be
\nabla_{ z}^{(N+1)}f_1 = \nabla_{ \bar z}^{(N-1)}f_2 = \nabla_{ z}^{(N-1)}f_0^H = \nabla_{ \bar z}^{(N+1)} f_0^L = 0\,. \lb{grstSP}
\ee
One immediately observes that, for any choice of $N\geq 0$, the first of eqs. \p{grstSP} has no normalizable  solution.
The other equations, depending on the value of $N$, yield the following non-trivial ground-state wave functions.
\begin{itemize}

\item For $N=0\,$,   one has two normalizable singlet ground states:
\be\label{N=0}
f_0^{H,0} (z,{\bar z}) ={ f_0^{H,0}\over 1 + {\bar z}z}, \ \ f_0^{L,0} (z,{\bar z}) ={ f_0^L\over 1 + {\bar z}z}.
\ee
where $f_0^{H,0}$ and $f_0^{L,0}$ are constants. Thus in this case the ground states are $SU(2)$ singlets.

\item For $N={1 \over 2}\,$,  one has normalizable doublet ground states:
\be\label{Nonehalf}
 f_0^{L,0} (z,{\bar z}) ={ A + B z \over (1 + {\bar z}z)^{3\over 2}},
\ee
the constants $A$ and $B$ thus forming spin $1/2$ multiplet of $SU(2)$.

\item For $ N\geqslant 1\,$, one has the following set of the ground states:
\be\label{Nbiggerone}
f_2^0 (z,{\bar z}) ={ f_2^0(z)\over (1 + {\bar z}z)^{N- 1}},\ N_{ {max}} = 2(N -1), \ \  f_0^{L,0} (z,{\bar z})
={ f_0^{L,0}(z)\over (1 + {\bar z}z)^{N +1}},\ N_{max} = 2N.
\ee
Here, $f_2^0(z)$ and  $f_0^{L,0}(z)$ are polynomials in $z$ of the maximum degree $N_{max}$, thus implying that the ground
states carry spins $N-1$ and $N$ (the coefficients of the $z$ monomials are just the components of the corresponding
$SU(2)$ multiplets, like in \p{Nonehalf}).\footnote{Under $SU(2)$, the polynomial $f(z)$ of the maximal degree $N_{max}$
transform as $\delta f(z) = N_{max}\,\bar\varepsilon z\, f(z) - \delta z f'(z)\,$. This generic transformation law agrees with the
laws \p{su2wf}.}
\end{itemize}

In accord with the general relations \p{N2susyVAC}, all ground states are singlets under the ${\cal N}=2$ SUSY transformations,
which can be directly checked using eqs. \p{QL}, \p{QH} adapted to the case at hand.
\vspace{0.2cm}

\noindent{\bf Higher LL states}
\vspace{0.2cm}

The non-zero eigenvalues for supersymmetric partners, $f_1$ and $f_0^L\,$, go by the standard pattern, and for $ N\geqslant 0$ one has
\bea
E_{\ell} = \ell (\ell + 2N + 1), \quad \ell = 1, 2 \ldots \,, \lb{LL1}
\eea
\bea
&& f_ 1^{1}= {\tilde f}_1^{1}, \quad  f_ 1^{\ell} = \nabla _ z^{(N+ 3)}\cdots \nabla _ z^{(N+2\ell-1)}{\tilde f}_1^{\ell}, \;\; \ell > 1\,,
\lb{f1} \lb{LL2} \\
&& \nabla_{\bar z}^{(N + 1)}{\tilde f}_1^{\ell} = 0 \; \Rightarrow \; {\tilde f}_1^{\ell}
= {{\tilde f}_1^{\ell}(z) \over (1 + {\bar z} z)^{N+1}}, \quad
f_0^{L,\ell}=  \nabla_z^{(N+ 1)}\hat{f}_1^{\ell}\,, \ \ell \geq 1\,,\lb{f0L}
\eea
where $\hat{f}_1{}^{\ell}(z, \bar z)$ is expressed in terms of an analytic function $\hat{\tilde f}{}_1^{\ell}(z)$ in precisely the same way
as ${f}_1^{\ell}(z, \bar z)$ is in terms of  ${\tilde f}_1^{\ell}(z)\,$, in \p{LL2}.
>From the computation of the norm of $f_ 1^{\ell}$ and $\hat{f}{}_1^{\ell}$, it follows that
the polynomials ${\tilde f}_1^{\ell}(z)$ and  $\hat{\tilde{f}}{}_1^{\ell}(z)$ have the maximum degree
$N_{ {max}} = 2(N+\ell)$. The convergence of the norm of $ f_0^{L,\ell}$ is then guaranteed by that of the norm of  $\hat{f}_ 1^{\ell}$,
upon performing an integration by parts. Thus the LL states with $\ell \geq 1$ are spanned by two independent $SU(2)$
multiplets of  spin $N + \ell$ encoded in  the wave functions ${\tilde f}_1^{\ell}(z)$ and $\hat{\tilde{f}}{}_1^{\ell}(z)\,$.
This additional two-fold degeneracy of the spectrum is of course a consequence of ${\cal N}=2$ supersymmetry which
transforms ${\tilde f}_1^{\ell}(z)$ and $\hat{\tilde{f}}{}_1^{\ell}(z)\,$ into each other and commutes with $SU(2)$.

This sequence of eigenvectors and eigenvalues can be extended to include the ground (LLL) states for $f_0^{L,0}$
from \p{N=0} - \p{Nbiggerone} and  correspondingly, to admit $\ell = 0$ in the eigenvalues \p{LL1}.
Since $f_0^{L,0}$ is a singlet of ${\cal N}=2$ supersymmetry, no two-fold degeneracy comes out at $\ell =0\,$.
The completed set of eigenvalues is given by
\be\label{Eell'}
E_{\ell'}^L = \ell' (\ell' + 2N + 1), \quad  \ell' = 0, 1 \ldots \,.
\ee

Now we shall focus on the second ${\cal N}=2$ multiplet of wave functions.
The non-zero eigenvalues of supersymmetric partners $f_2$ and  $f_0^H$ must be split according
to  $0\leqslant N < 1$ and $N\geqslant 1$, as implied by the eigenvalue equation for $f_2$, which demands
that for $0\leqslant N < 1$   we should work on the subspace of anti-analytic functions.

For $N\geqslant 1$ one has
\bea
&&\  \  \  \  \  \  \ E_{\ell} = \ell  (\ell  + 2N -1), \quad  \ell = 1 \ldots \,, \lb{spectre2} \\
 && f_ 2^{\ell} = \nabla _ { z}^{(N+1)}\cdots \nabla _ { z}^{(N+2\ell -1)}{\tilde f}_2^{\ell}, \quad
   \nabla_{\bar z}^{(N - 1)}{\tilde f}_2^{\ell} = 0 \; \Rightarrow \;  \\
   &&{\tilde f}_2^{\ell} = {{\tilde f}_2^{\ell}({ z}) \over (1 + {\bar z} z)^{N-1}}\,, \lb{spectre3}
    \ \  f_0^{H,\ell} =  \nabla_{\bar z}^{(N- 1)}\hat{f}_2^{\ell}\,,
\eea
where $\hat{f}_2^\ell$ is expressed through an analytic function $\hat{\tilde{f}}_2^\ell$ in the same way as
${f}_2^\ell$ through ${\tilde{f}}_2^\ell\,$. From the computation of the norm of $f_ 2^{\ell}$ and $\hat{f}_2^{\ell}$,
it follows that   the polynomials ${\tilde f}_2^{\ell}(z)$ and $\hat{\tilde{f}}_2^\ell$
have the maximal degree $N_{ {max}} = 2(N+\ell -1)$. The convergence of the norm of $ f_0^{H,\ell}$ is then guaranteed
by that of the norm of  $\hat{f}_ 2^{\ell}\,$. Thus, like in the previous case, we observe two-fold degeneracy of the energy
spectrum due to ${\cal N}=2$ supersymmetry, having two irreducible $SU(2)$ multiplets with spin $N+ \ell -1\,$.
Extending the range of $\ell$ to include $0$ for the ground state vectors
$f_2^0(z,{\bar z})$ from \p{Nbiggerone}, one eventually obtains the full second sequence of eigenvectors corresponding to
\be
E_{\ell}^H = \ell  (\ell  + 2N -1),  \quad \ell = 0,1 \ldots \,.
\ee
Once again, no two-fold degeneracy occurs at $\ell = 0$ because $f_2^0(z,{\bar z})$ are singlets of ${\cal N}=2$ supersymmetry.

To summarize the above discussion, for $N\geqslant 1$ the eigenvalues and eigenfunctions are split into two sequences corresponding
to two super monopole systems, one  with the charge $2N$ and the other with the charge  $2(N-1)$. The first sequence extends to the
entire range of $N \geqslant 0\,$.

It remains to analyze the case $0\leqslant N < 1\,$ for the multiplet $(f_2, f_0^H)$. We have the following non-zero eigenvalues,
\be
E_{\ell} = (\ell + 1 ) (\ell  - 2N + 2),  \quad \ell = 0, 1 \ldots \,. \lb{sequence4}
\ee
\bea
&& f_ 2^{0}= {\tilde f}_2^{0}; \quad  f_ 2^{\ell} = \nabla _ {\bar z}^{(N- 3)}\cdots \nabla _ {\bar z}^{(N-2\ell-1)}{\tilde f}_2^{\ell}, \ \
\; \ell > 0\,; \nn
&& \nabla_{ z}^{(N - 1)}{\tilde f}_2^{\ell} = 0 \; \Rightarrow \;
{\tilde f}_2^{\ell} = {{\tilde f}_2^{\ell}({\bar z}) \over (1 + {\bar z} z)^{1-N}}\,, \quad f_0^{H,\ell}=  \nabla_{\bar z}^{(N- 1)}\hat{f}_2^{\ell},
\eea
where $\hat{f}_2^{\ell}$ is related to an anti-analytic function $\hat{\tilde{f}}_2^{\ell}$ as ${f}_2^{\ell}$ is to
${\tilde{f}}_2^{\ell}\,$. From the computation of the norms of $f_ 2^{\ell}$ and $\hat{f}_2^{\ell}$, it follows that
the polynomials  ${\tilde f}_2^{\ell}({\bar z})$ and $\hat{\tilde{f}}_2^{\ell}(\bar z)$ have the maximal degree
$N_{ {max}} = 2(-N+\ell +1)$ and, hence, encompass two independent $SU(2)$ multiplets with spin $1-N+\ell$, revealing
the same two-fold degeneracy as in the previous cases. The convergence of the norm of $ f_0^{H,\ell}$ is then guaranteed
by that of the norm of  $\hat{f}_ 2^{\ell}\,$.

For $N = 0$, one can make the shift $\ell' = \ell + 1$ and append  the value $\ell' = 0$ associated
with the ground-state function $f_0^{H,0}$ from \p{N=0}, obtaining in this way the completed set of eigenvalues as
\be
E_{\ell'}^{(N = 0)} = \ell'  (\ell' + 1),  \ \ \ \ell' = 0, 1 \ldots \,.
\ee
This set for $\ell' > 0$ is clearly degenerate with the corresponding $N=0$ set from \p{Eell'}. Therefore, in this case
the system acquires an extra degeneracy: excited levels built on  the corresponding ${\cal N}=2$ singlet ground states
possess the same energy. So in this case the system reveals a four-fold
degeneracy (like in the superplane Landau model \cite{IMT3,CIMT}).

For $N = {1 \over 2}$, there is no match for the singlet ground state \p{Nonehalf} in the above sequence,
so in this sector ${\cal N}=2$ supersymmetry
appears as spontaneously broken, even though for the whole system it is not, because for the other
supermultiplet $f_1, f_0^L$,  in the range $N\geqslant 0$, there is always an ${\cal N}=2$ supersymmetric singlet ground state.

Finally, let us note that, should we have chosen $N \leqslant 0$, we would expect that the role of $f_1, f_0^L$ and analyticity
will be replaced by   $f_2, f_0^H$,  and anti-analyticity (and vice-versa).

\section{Summary and outlook}

We have shown that the world-line ${\cal N}=2$ supersymmetry is strong enough to define unambigously a rather general family of quantum models.
Moreover, the component Hamiltonians so obtained  have a structure similar to those
investigated in \cite{FV}, so similar conclusions emerge. The naive definition of the inner product has been easily
modified so that the illusory ghosts were resurrected as proper, positive-normed states in the redefined models.

The fact that we have used systems for which kinetic terms of the odd variables were quadratic in time
derivatives has led to general wave functions containing reducible representations of supersymmetry,
as can be seen by contemplating \p{Rep1}, \p{q1}, and \p{q2}. Should we have used standard
kinetic terms that are linear in time derivatives of  the anticommuting variables, the general two-component
wave functions would have carried irreducible representations of supersymmetry. One can imagine that such an approach,
should it hold for other theories, may well complement the more familiar group theoretic approach to unification.

In a previous paper \cite{CIMT} we exhibited a target space model which had a hidden ${\cal N} = 2$ supersymmetry.
In this paper we found target superspaces with built-in ${\cal N} = 2$ supersymmetry, and corresponding connections
(see \p{inversemetric'}, and \p{u1conn}) whose geometries have certain interesting features (see  footnote 4).

 For the special case of constant Gauss curvature and constant magnetic fields one can connect our results with
 analogous ones of previous investigations.
 Standard ${\cal N} = 2$ supersymmetric actions which are linear in time derivatives of the anticommuting variables, that is
 actions for a particle on a sphere in the field of a monopole, have been treated in a number
 of publications \cite{D'Hoker:1983xu}-\cite{Hong:2005ce}. The wave function has  two components,
 each of them belonging to a representation of $SU(2)$ and transforming into each other under ${\cal N}=2$ supersymmetry.
 The generic wave function (with SUSY singlets excluded) contains one SUSY lowest  weight component which
 can belong to different energy eigenvalues.  It can be seen from \cite{Hong:2005ce} that
 for $N= {1 \over 2}$ the supersymmetry is spontaneously broken, as the zero-energy state is not allowed for this case.

 In our non-minimal model the generic four-component wave function contains two SUSY lowest weights
 of different charges $2N$ and $2(N -1)$. If  $N\geqslant 1$ there are two families of zero-energy
 solutions annihilated by the supercharges (see formula \p{Nbiggerone}). For the case of $N = 0$, the system
 acquires an additional degeneracy. An interesting turn occurs for
 $N= {1 \over 2}$, where one of the ground states corresponding to the charge $-1$
 is not admissible like in the generic ${\cal N}=2$ supersymmetric models (see \cite{Hong:2005ce}), but the other ground
 state is admissible keeping the whole supersymmetry  unbroken. That is to say, in our non-minimal model
 supersymmetry is unbroken for any strength of the monopole.

Finally, we would like to emphasize two possible directions for further study. First, it is interesting to inquire
whether the general model \p{a} admits some non-trivial super-~isometries for special choices of the potentials $K$ and $V$,
like its superplane propotype \p{7}, which is known to respect $ISU(1|1)$ super-isometry \cite{IMT3}. Another task is
to construct super Landau models with ${\cal N}=4$ and higher ${\cal N}$ world-line supersymmetries. Such models are not known
even in the planar limit. They could bear a close relation to the Landau-type models on higher  dimensional spaces, e.g.
to the  LM on $R^4$ \cite{ElPo}.

\bigskip
\section*{Acknowledgements}
We benefited from useful discussions with  A. Schwimmer,  A. Smilga, and J. Sonnenschein.
We are specially indebted to Paul K. Townsend for collaborating at the early stages of this study.
T.C. and L.M. acknowledge partial support
from the National Science Foundation Award 0855386.  L.M.  thanks the Directorate of BLTP, JINR (Dubna)
and Department of Physics of the Weizmann Institute of Science for the kind hospitality extended
to him in the course of this work. Parts of these investigations
were presented by L.M. in the conference talks at ``Supersymmetries
and Quantum Symmetries - SQS'09'' July 29 - August 3, 2009, Dubna, Russia,
``Miami 2009'', 15 - 20 December 2009 Fort Lauderdale, Florida and
at ``Joint Seminars in Theoretical High Energy Physics
Hebrew University, Tel-Aviv University, Weizmann Institute at Neve Shalom/Wahat al-Salam''.
E.I. acknowledges partial support from the RFBR grants 09-02-01209, 09-02-91349 and 09-01-93107.

\end{document}